\documentclass[a4paper,fleqn,usenatbib]{mnras}
\usepackage[T1]{fontenc}
\usepackage{ae,aecompl}

\usepackage{subfig}
\usepackage{graphicx}
\usepackage{caption}
\usepackage{float}
\usepackage{amsmath}
\usepackage{amssymb}
\usepackage{footnote}

\title[$\left\langle \Pi \right\rangle$ of extragalactic dusty sources in \textit{Planck}]
{Statistics of the fractional polarisation of extragalactic dusty sources in \textit{Planck} HFI maps}

\author[L. Bonavera]{
L. Bonavera,$^{1}$\thanks{E-mail: laurabonavera@gmail.com}
J. Gonz\'{a}lez-Nuevo,$^{1}$
B. De Marco,$^{2}$
F. Arg\"ueso,$^{3}$
L.  Toffolatti$^{1,4}$
\\
$^{1}$Departamento de F\'{i}sica, Universidad de Oviedo, C. Federico Garc\'{i}a Lorca 18, 33007 Oviedo, Spain\\
$^{2}$Nicolaus Copernicus Astronomical Center, Polish Academy of Sciences, Bartycka 18, PL-00-716 Warsaw, Poland\\
$^{3}$Departamento de Matem\'aticas, Universidad de Oviedo, C. Federico Garc\'{i}a Lorca 18, 33007 Oviedo, Spain\\
$^{4}$INAF/IASF Bologna, Via Gobetti 101, Bologna, Italy\\
}

\date{Accepted XXX. Received YYY; in original form ZZZ}

\pubyear{2016}

\begin{document}
\label{firstpage}
\pagerange{\pageref{firstpage}--\pageref{lastpage}}
\maketitle

\begin{abstract}
We estimate the average fractional polarisation at 143, 217 and 353 GHz of a sample of 4697 extragalactic dusty sources by applying stacking technique. The sample is selected from the second version of the \textit{Planck} Catalogue of Compact Sources at 857 GHz, avoiding the region inside the {\it Planck} Galactic mask ($f_{sky}$ $\sim 60$ {\it per cent}).
We recover values for the mean fractional polarisation at 217 and 353 GHz of ($3.10 \pm 0.75$) {\it per cent} and ($3.65 \pm 0.66$) {\it per cent}, respectively, whereas at 143 GHz we give a tentative value of ($3.52 \pm 2.48$) {\it per cent}. We discuss the possible origin of the measured polarisation, comparing our new estimates with those previously obtained from a sample of radio sources.

We test different distribution functions and we conclude that the fractional polarisation of dusty sources is well described by a log-normal distribution, as determined in the radio band studies.
For this distribution we estimate $\mu_{217 GHz}=0.3 \pm 0.5$ (that would correspond to a median fractional polarisation of $\Pi_{med}=(1.3 \pm 0.7)$ {\it per cent}) and $\mu_{353 GHz}=0.7 \pm 0.4$ ($\Pi_{med}=(2.0 \pm 0.8)$ {\it per cent}), $\sigma_{217 GHz}=1.3 \pm 0.2$ and $\sigma_{353 GHz}=1.1 \pm 0.2$.
With these values we estimate the source number counts in polarisation and the contribution given by these sources to the CMB B-mode angular power spectrum at 217, 353, 600 and 800 GHz. We conclude that extragalactic dusty sources might be an important contaminant for the primordial B-mode at frequencies $ > 217$ GHz.

\end{abstract}



\section{Introduction}

The polarisation properties of extragalactic dusty sources, i.e. sources dominated by thermal dust emission, at high frequencies ($> 100 GHz$) are still very poorly constrained (if not at all) by observations. For M82 a value for the fractional polarisation $\Pi$ of 0.4 per cent at $850 \mu m$ has been measured by \cite{GRE02} with SCUPOL. 
The polarisation data from {\it Planck} provide the first all-sky map of the polarised dust emission of our Galaxy \citep{PIPXIX}. From the  {\it Planck} dust polarisation maps \cite{DEZ17} has recently inferred an estimate of the fractional polarisation of our Galaxy: they found an average polarisation degree of $2.7$ per cent for the Stokes Q parameter by integrating over a spatial band centered in the galactic plane and with $20^\circ$ of width in latitude. This translates into a mean fractional polarisation, $\left\langle \Pi \right\rangle$, of $1.4$ per cent if all galaxies were polarised like the Milky Way, by considering all the possible orientations of galactic disks with respect to the sky sphere.

The intrinsic Cosmic Microwave Background (CMB) polarised signal is very weak and primordial B-modes are yet to be discovered. The current upper limit on the tensor-to-scalar ratio from the BICEP2 and Keck Array experiments is $r \sim 0.07$ \citep{BIC15,KEC16}. Moreover, current measurements of the Galactic foreground emission \citep{CHO15,PIP30} imply that primordial B-modes would be sub-dominant with respect to foregrounds on all angular scales and over all observational frequencies in the microwave regime. For this reason, the detection of B-modes should be regarded as a component separation problem \citep[see][and references therein]{REM17}. Therefore, future CMB all-sky surveys in polarisation \citep[e.g., the proposed European Space Agency, ESA, Cosmic ORigin Explorer, CORE, mission, see][]{CORE} -- with the capability to reach tensor-to-scalar ratios down to $r\sim 0.01$ -- will need a much better characterization of the contaminating signal due to the (dominant) Galactic diffuse emission but also to compact extragalactic sources \citep[see][for comprehensive reviews on this subject]{DEZ15,DEZ17}.

In general, the contamination due to compact source to the polarisation signal of the CMB does not affect large angular scales (near the reionization peak), but can play an important role at the intermediate and small angular scales of the lensing-induced B-mode signal \citep{CUR13}. Indeed, dusty galaxies  are expected to be the dominant foreground for $ r \sim 0.001 $ at small angular scales \citep[$l > 47$][]{CUR13,REM17}, once delensing has been applied to the data. 

The polarisation properties of populations of faint compact sources at mm wavelengths are poorly constrained. 
The currently available all-sky catalogues of compact sources at mm/sub-mm wavelengths are still limited to the shallow surveys provided by the Wilkinson Microwave Anisotropy Probe (WMAP) \citep{BEN03} and the ESA {\it Planck} \citep{PLA15gen} missions. In the second, updated, version of the \textit{Planck} Catalogue of Compact Sources (PCCS2) \citep{PLA15pccs2} the number of detected compact sources in polarisation is very low (i.e. 25 at 143 GHz, 11 at 217 GHz and just one at 353 GHz at the $99.99$ per cent confidence level) and, thus, their polarisation properties are poorly characterized. Moreover, it is obvious that only compact -- either Galactic or extragalactic -- sources with a high $\Pi$ can be detected and, thus, the statistical characterization of the underlying population will be biased towards these highly polarised objects.

In order to gather information about the statistical properties of the fractional polarisation of compact sources in the microwave band, it is useful to exploit the full information embedded in CMB all-sky polarisation maps by applying stacking techniques, i.e., by co-adding the signal from many weak or undetected objects to obtain a statistical detection. 
Recently, Bonavera et al. \citep[2017;][]{BON17} have applied stacking to {\it Planck} maps to recover $\left\langle \Pi \right\rangle$ from 30 to 353 GHz of a primary sample of 1560 compact radio sources. The sources were selected in the PCCS2 at 30 GHz and were divided according to their location, i.e. whether located outside or inside the Galactic region of the sky defined by the {\it Planck} GAL60 mask and the region around the Magellanic Clouds. They found that $\left\langle \Pi \right\rangle$ is approximately constant with frequency in both samples, with a weighted mean value for all the {\it Planck} channels of $3.08$ per cent outside and $3.54$ per cent inside the Galactic mask, respectively. In the extragalactic region they estimated the parameters ($\mu$ and $\sigma$) for the log-normal distribution of $\Pi$, finding a weighted mean value of 1.0 for $\sigma$ and 0.7 for $\mu$, that would imply a weighted median value for $\Pi$ of 1.9 per cent. 

Here we extend the stacking approach to a sample of dusty galaxies, with the goal of estimating their fractional polarisation,  source number counts and their contribution to the primordial CMB polarised signal.
The outline of the paper is as follows: in Section \ref{sec:met} we discuss the methods adopted for selecting the sky patches, for defining the sub-samples we are going to analyse and for determining their mean fractional polarisation; in Section \ref{sec:results} we present our results and in Section \ref{sec:conc} our conclusions.

\section{Methods}
\label{sec:met} 

\subsection{Data}
\label{sec:data} 

\begin{figure*}
 \centering
 \includegraphics[width=16.0cm]{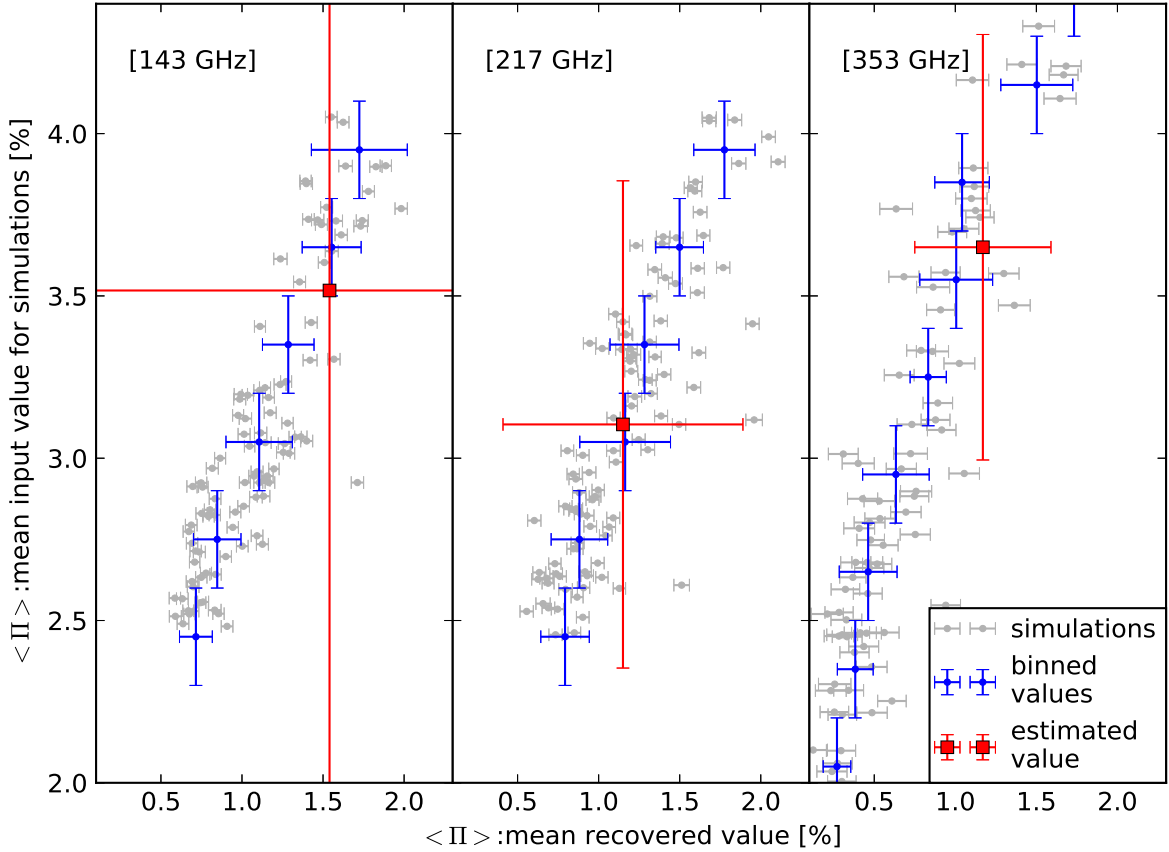}
 \caption{Results obtained outside the adopted mask ({\it extragalactic region} of the sky) from 143 to 353 GHz. The grey points are obtained with each individual simulation (using a log-normal distribution for the mean fractional polarisation): on the y-axis we plot the mean input $\left\langle \Pi \right\rangle $ value for simulations and the x-axis is the value recovered with stacking for different values of $\mu$ and $\sigma$, as described in the text. For all the panels, the linear interpolation of these points gives us the correction for the noise bias that has to be applied to the observed values (red squares). The blue points are obtained by averaging over the simulations points with a binning step of 0.3 in the y-axis. }
 \label{fig:ex_res}
\end{figure*}

\begin{figure*}
 \centering
 \includegraphics[width=16.0cm]{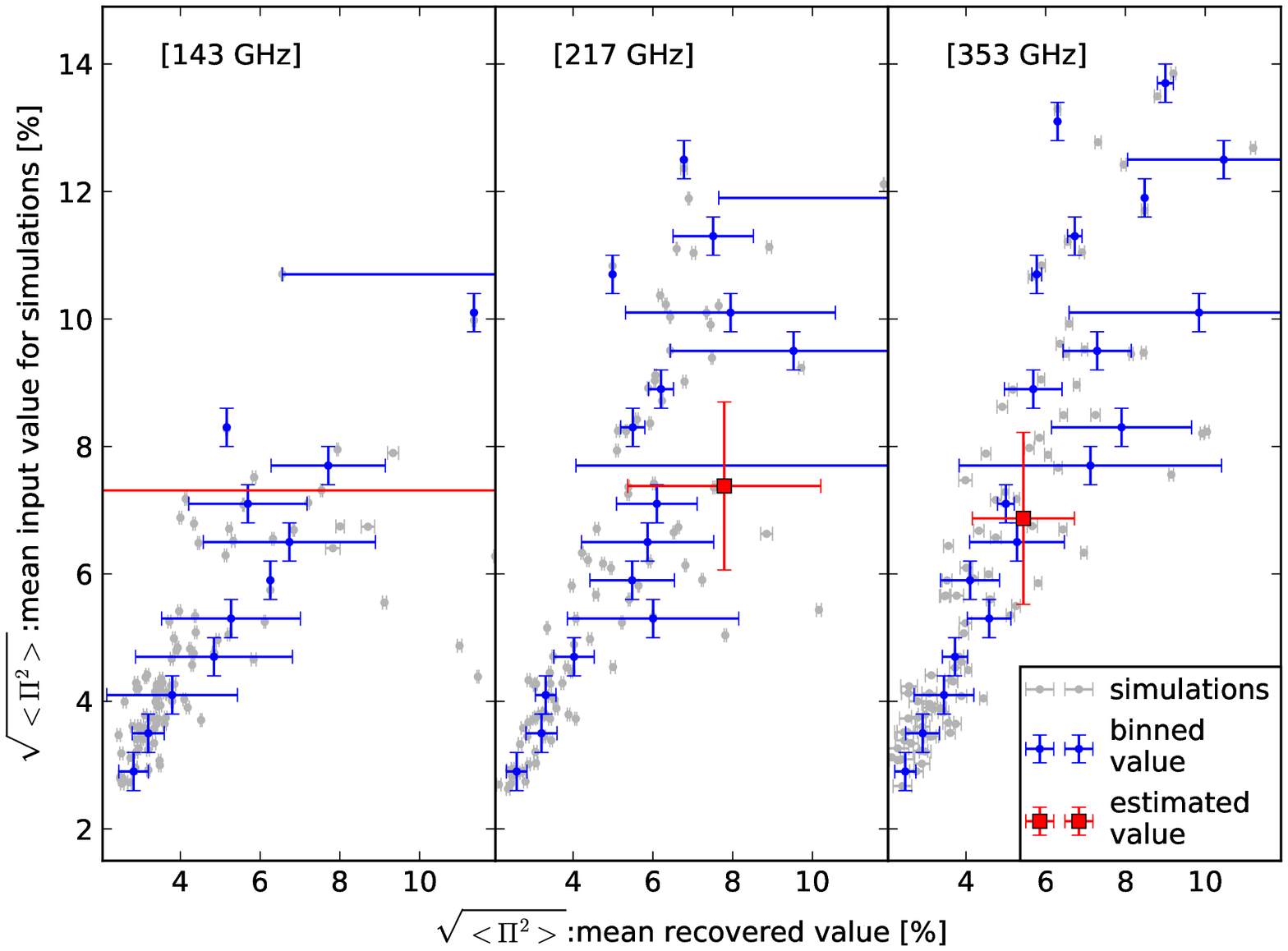}
 \caption{Results obtained for $\sqrt{\left\langle \Pi^2 \right\rangle} $  outside the adopted mask ({\it extragalactic region} of the sky) from 30 to 353 GHz. The grey points are obtained with each individual simulation (using a log-normal distribution for the mean fractional polarisation): the y-axis is the mean input $\sqrt{\left\langle \Pi^2 \right\rangle} $  for simulations and the x-axis is the value recovered with stacking for different values of $\mu$ and $\sigma$, as described in the text. The linear interpolation of these points gives us the correction for the noise bias that has to be applied to the observed values (red squares). The blue points are obtained by averaging over the simulations points with a binning step of 0.3 in the y-axis.}
 \label{fig:ex_res_pi2}
\end{figure*}

\begin{figure}
 \centering
 \includegraphics[width=\columnwidth]{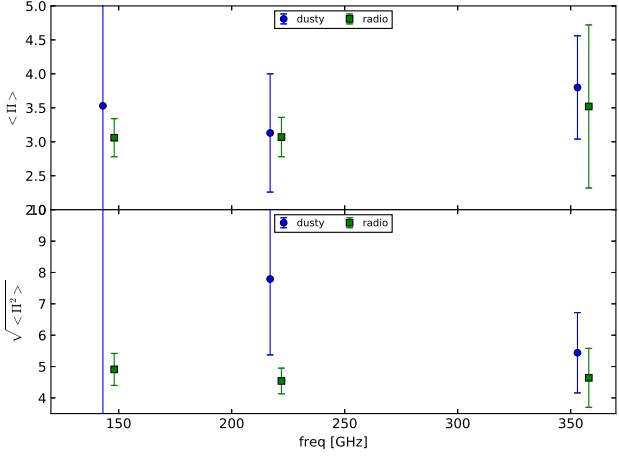}
 \caption{$\left\langle \Pi \right\rangle$ (top) and $\sqrt{\left\langle \Pi^2 \right\rangle}$ (bottom) for each frequency outside the Galactic mask with $f_{sky}$=60 per cent and the Magellanic Clouds. The blue circles are obtained for the current sample, the green squares refer to the sample discussed in \protect\cite{BON17}.}
 \label{fig:polfrac}
\end{figure}

\begin{figure}
 \centering
 \includegraphics[width=\columnwidth]{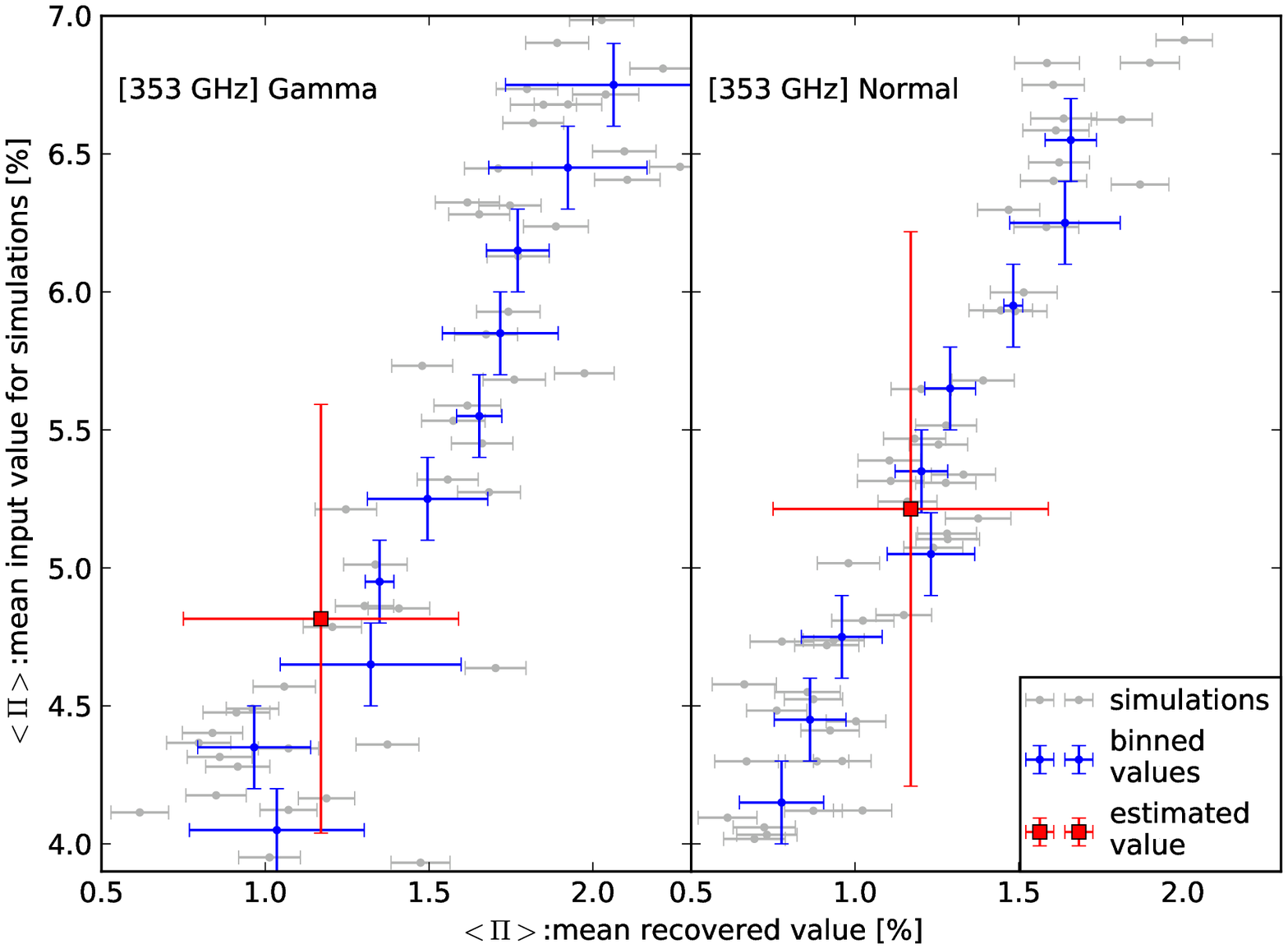}
 \caption{Results obtained outside the adopted mask ({\it extragalactic region} of the sky) for the 353 GHz case. The grey points are obtained with each individual simulation: on the y-axis we plot the mean input $\left\langle \Pi \right\rangle $ value for simulations and the x-axis is the value recovered with stacking for different values of the mean for the Normal (left) and gamma (right) distributions. Linear interpolation and blue points as in Figs. \ref{fig:ex_res_cfr} and \ref{fig:ex_res_cfr_pi2}.}
 \label{fig:ex_res_cfr}
\end{figure}

\begin{figure}
 \centering
 \includegraphics[width=\columnwidth]{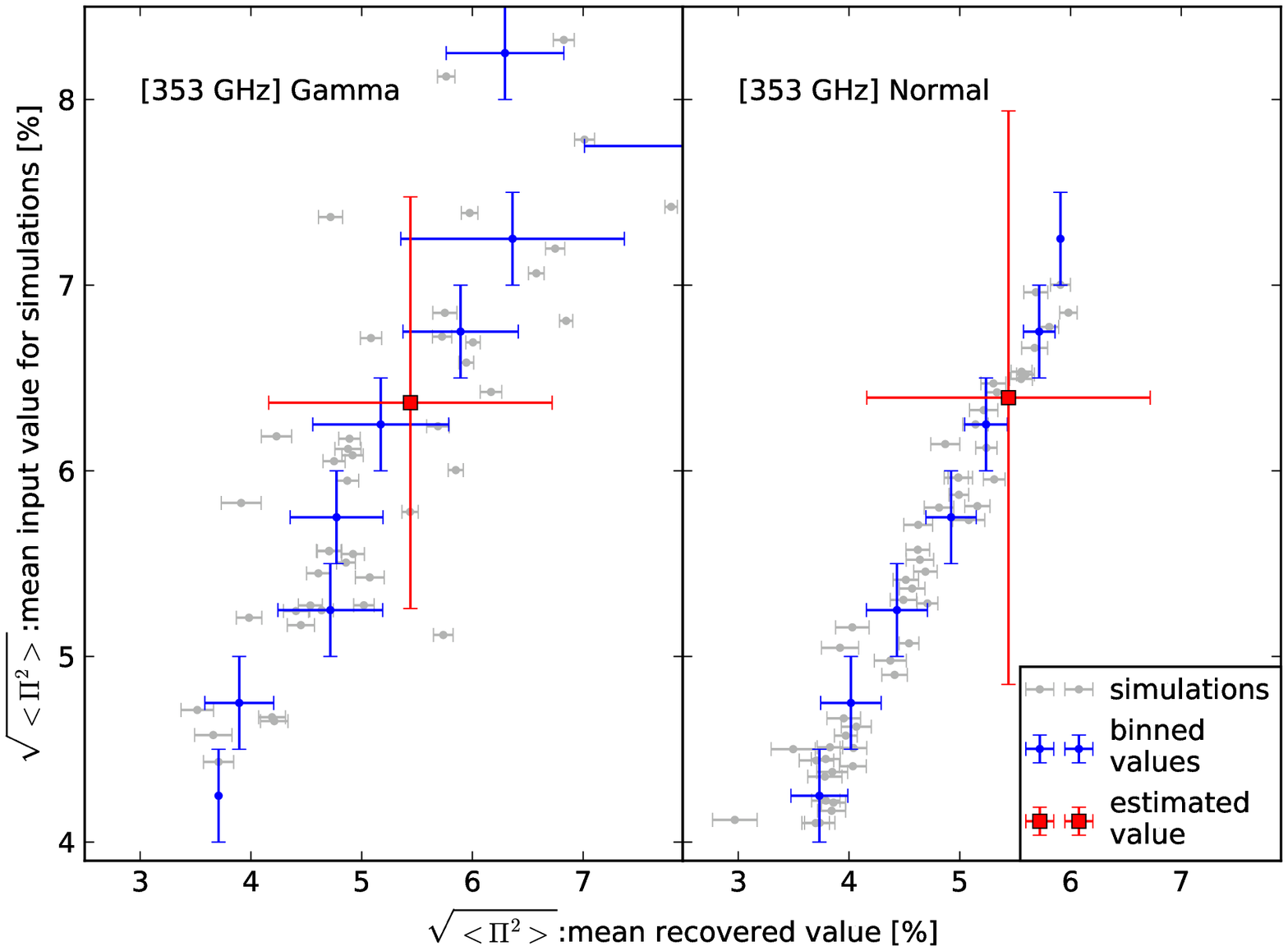}
 \caption{Results obtained outside the adopted mask ({\it extragalactic region} of the sky) for the 353 GHz case. The grey points are obtained with each individual simulation: on the y-axis we plot the mean input $\sqrt{\left\langle \Pi^2 \right\rangle} $ value for simulations and the x-axis is the value recovered with stacking for different values of the mean for the Normal (left) and gamma (right) distributions. Linear interpolation and blue points as in Figs. \ref{fig:ex_res_cfr} and \ref{fig:ex_res_cfr_pi2}.}
 \label{fig:ex_res_cfr_pi2}
\end{figure}

Our sample is based on the PCCS2 \citep{PLA15pccs2} catalogue\footnote{Based on observations obtained with {\it Planck} (http://www.esa.int/Planck), an ESA science mission with instruments and contributions directly funded by ESA Member States, NASA, and Canada. Available at http://pla.esac.esa.int/pla/\#home (\textit{Planck} Legacy Archive).} at 857 GHz that contains 4891 sources with a flux density at the 90 per cent completeness level  of 791 mJy in total intensity. In order to avoid radio sources in our sample, we exclude those sources that have a couterpart in the 30 GHz PCCS2 catalogue with a 33 arcmin search radius (151 sources). Of these 4740 sources, for consistency with what was done in the radio bands by \cite{BON17}, we consider for our analysis only those sources lying in the region outside the {\it Planck} galactic mask that leaves the 60 per cent of the sky unmasked and we also exclude the 5 $deg$ radius region around the position of the Large Magellanic Cloud and the 3 $deg$ radius region around the Small Magellanic Cloud. 

To check for further possible radio sources among these 4706 sources, we search in the PCCS2 217 GHz catalogue for sources in the PCCS2 143 GHz within a 7.1 arcmin search radius from the center of the sources at 217 GHz. The resulting 944 sources are then cross-matched with our sample (the one selected at 857 GHz and without the sources in the 30 GHz channel). We find only 18 sources in common. Of these, only 9 have a radio spectrum ($\alpha < 1$), but their flux densities are too faint (the maximum flux density is lower than 400 mJy) to be detected at 857 GHz. In any case, we remove them to minimize any potential contamination when performing stacking at the lower frequencies. Our final sample consists of 4697 sources.

\begin{table*}
 \centering
\begin{tabular}{ l | c | c | c | c  | c  | c | c | c | c | c | c | c | c || r } 
\hline
&  \multicolumn{12}{c}{Extragalactic region} \\
 &  \multicolumn{2}{c}{Uncorrected} & \multicolumn{2}{c}{Corrected} &  \multicolumn{2}{c}{Uncorrected} & \multicolumn{2}{c}{Corrected} & \multicolumn{6}{c}{Log-normal parameters} \\
freq &  $\left\langle \Pi \right\rangle$ & error &  $\left\langle \Pi \right\rangle$ & error & $ \sqrt{\left\langle \Pi ^2 \right\rangle}$ & error & $ \sqrt{\left\langle \Pi ^2 \right\rangle}$ & error & $\mu$ & error & $\sigma$ & error & $\Pi_m$ & error \\
$[GHz]$  &\% & \% & \% & \%  & \% & \%  & \% & \% & & &  & & \% & \%\\
\hline
143 & 1.54 & 2.42 & 3.52 & 2.48 & 12.56 & 12.14 & 7.31 & 3.89 & 0.53  & 1.51 & 1.21 & 0.73 & 1.6 & 2.5\\ 
217 & 1.15 & 0.74 & 3.10 & 0.75 & 7.79 & 2.42 & 7.38 & 1.32 & 0.26  & 0.52 & 1.32 & 0.23 & 1.3 & 0.7\\ 
353 & 1.17 & 0.42 & 3.65 & 0.66 & 5.44 & 1.28 & 6.87 & 1.35 & 0.66  & 0.41 & 1.12 & 0.24 & 2.0 & 0.8\\ 
\hline
\end{tabular} 
 \caption{From left to right: frequency, mean fractional polarisation with r.m.s. errors uncorrected and corrected for the noise bias, square root of the mean quadratic fractional polarisation with 1$\sigma$ errors uncorrected and corrected for the noise bias, $\mu$ and $\sigma$ parameters of the log-normal function characterizing the mean fractional polarisation distributions (see text) and their 1$\sigma$ errors and median fractional polarisation computed as $\Pi_m$ and its error. The results are for the case outside the \textit{Planck} Galactic mask with $f_{\rm sky}$ = 60 per cent and excluding the regions around the Magellanic Clouds as described in the text.}
   \label{tab:polfrac_ext}
 \end{table*}

\subsection{Stacking}
\label{sec:stacking} 

In this work we apply the same methodology discussed in \cite{BON17}: we use stacking \citep[see][and references therein]{DOL06,MAR09,BET12} to reduce the noise/background, since it is expected to fluctuate around the mean with positive and negative values, and enhance the signal we want to study, correspondingly.

In our case we want to perform statistical estimates of polarisation with {\it Planck}. It should be noticed that our target sources are all detected by {\it Planck} in total intensity at 857 GHz, but they are not necessarily detected in the lower frequency channels. 
At 857 and 545 GHz {\it Planck} instruments are not sensitive to polarisation. For this reason we compute $\left\langle \Pi \right\rangle$ of our source population at 143, 217 and 353 GHz. Due to the rising spectra with frequency of dusty sources, we do not expect to have a measurable signal either in total intensity or in total polarisation at frequencies lower than 143 GHz.

To be consistent with \cite{BON17}, we perform stacking selecting the same small patch of 63 x 63 pixels (with a pixel size of 1.72') around each source position. We then add up all the patches to obtain the total flux density. To reduce the instrumental noise (a second order effect) we convolve the resulting patch  with a Normal filter whose $\sigma_{\rm filter}$ is given by $\sigma_{\rm beam}/2$, where $\sigma_{\rm beam}$ is $FWHM_{instrument}/2\sqrt{2ln2}$.

Following \cite{BON17}, we remove plausible contaminants to our stacked measurements that result in a strong background in the final stacked image. We subtract the mean of the background computed in the external region of the final patch (3$\sigma_{\rm beam}$ away from the patch center) from the total flux density. From these residual maps we then compute the total flux densities in total intensity $S$ and polarisation $P_0$. We then obtain the averaged values $\left\langle P_0 \right\rangle$ and $\left\langle S \right\rangle$ over our sample.

Finally we compute $\left\langle \Pi \right\rangle = \left\langle P_0 \right\rangle / \left\langle S \right\rangle$. Its error is given by 
$ \sqrt{ \left( \left\langle P_0 \right\rangle / \left\langle S \right\rangle \right)^{2}  \cdot \left(  \sigma_{P_0}^2 /  \left\langle P_0 \right\rangle ^2 + \sigma_S ^2/  \left\langle S \right\rangle ^2 \right) }$,
 where $\sigma_{P_0}$ and $\sigma_S$ are the standard deviations for total intensity and polarisation computed in the external region of the stacked patches. 
We also compute the quantity $\sqrt{\left\langle \Pi^2 \right\rangle} = \sqrt{\left\langle P_0^2 \right\rangle / \left\langle S^2 \right\rangle}$ by applying the same methodology. It can be a useful quantity to define the parameters for the fractional polarisation distribution.
Its error is given by $\sqrt{\dfrac{1}{4 \left\langle P_0^2 \right\rangle \left\langle S^2 \right\rangle}  \left(   \sigma^2_{P_0^2} + \left\langle P_0^2 \right\rangle^2/ \left\langle S^2 \right\rangle^2 \sigma^2_{S^2} \right)  }   $.

Please note that most of our sources are not directly detectable and therefore we cannot estimate directly $\left\langle \Pi \right\rangle=\left\langle P_0 /S \right\rangle$ and $\left\langle \Pi^2 \right\rangle=\left\langle P_0^2 /S^2 \right\rangle$. For this reason, in our stacking procedure we decided to calculate $\left\langle \Pi \right\rangle=\left\langle P_0 \right\rangle / \left\langle S \right\rangle$ and $\left\langle \Pi^2 \right\rangle=\left\langle P_0^2 \right\rangle/\left\langle S^2 \right\rangle$ that are good approximations for $\left\langle \Pi \right\rangle=\left\langle P_0 /S \right\rangle$ and $\left\langle \Pi^2 \right\rangle=\left\langle P_0^2 /S^2 \right\rangle$, assuming that $\Pi$ and $S$ could be considered independent variables as for the case in the radio band \citep{BON17}. Besides, the residual errors introduced by these assumptions are much smaller than the noise bias (see \cite{BON17}  and Section \ref{sec:s_injection} for more details) and the bias subtraction methodology described in Section \ref{sec:s_injection} also corrects any residual deviation  from the theoretical value.

\subsection{Source injection}
\label{sec:s_injection} 

As described in ditail in \cite{BON17}, the construction of $P=\sqrt{Q^2+U^2}$ introduces a bias usually called noise bias which has to be taken into account in order to obtain correct measurements. To estimate this bias, we carry out simulations with injected sources. This allows us to correct for any inaccuracy introduced by our assumptions, as stated above. We injected simulated compact sources in our real maps, at random positions but avoiding the positions of real sources.

The sources in our sample are those detected by {\it Planck} at 857 GHz and they are mainly dusty local galaxies. They are simulated in total intensity at each frequency independently following \cite{NEG13}. The latter uses the {\it Planck} Early release Compact Source Catalogue \citep[ERCSC][]{ERCSC} to carefully determine the luminosity function, and therefore the source number counts, of local dusty galaxies at 857, 545 and 353 GHz.

The flux densities in polarisation are first simulated as in \cite{BON17} following a log-normal distribution for different values of $\left\langle \Pi \right\rangle$ in each simulation.
In order to estimate the noise-bias correction, we inject simulated sources in the real maps by assuming a uniform random polarisation angle. We compute Q and U for each source creating a simulated sources map starting with the simulated catalogue and convolve it with the FWHM of the instrument (different for each 
$\nu$) and add it to the real Q and U maps.
For each simulation we randomly vary both the location ($\mu$) and the scale ($\sigma$) parameters of the log-normal distribution. We then apply the same methodology described in Section \ref{sec:stacking} to obtain a recovered (biased) value. We repeat this procedure for at least 100 simulations at each frequency. The single theoretical simulated values, $\left\langle \Pi \right\rangle = \left\langle P_s/S_s \right\rangle$, and the recovered ones are shown as grey points. The blue points are simply the binned values, equally spaced in the y-axis.

By comparing the theoretical simulated $\left\langle \Pi \right\rangle = \left\langle P_s/S_s \right\rangle$ with the recovered values we are able to obtain the noise bias correction relationships shown as blue points in Fig. \ref{fig:ex_res}. Finally, we estimate our debiased measurements (red points) using a linear interpolation of the derived noise bias correction relationships.

Then, we also check how the results change when using a different distribution for the fractional polarisation to correct the biased values. We chose to study the cases for  a Normal and a gamma distribution. We adopt the same procedure as for the log-normal distribution, randomly vary the $\mu$ parameter for the Normal distribution (and assuming $\sigma=0.5$) and the {\it shape}, $k$, and {\it scale}, $\theta$, parameters for the gamma distributions. We perform at least 50 simulations for each case.

We also compute the correction to $\sqrt{\left\langle \Pi^2 \right\rangle}$ in the same way as for $\left\langle \Pi \right\rangle $. 

As in \cite{BON17} from the recovered values of $\left\langle \Pi \right\rangle$ and $\sqrt{\left\langle \Pi^2 \right\rangle}$ we can compute the parameters $\mu$ and $\sigma$ characterizing the log-normal distribution:
 \begin{equation}
 \mu=\ln\left( \dfrac{\left\langle \Pi \right\rangle^2}{\sqrt{\left\langle \Pi^2 \right\rangle}}  \right) 
 \end{equation}
  \begin{equation}
\sigma = \sqrt{\ln \left(  \dfrac{\left\langle \Pi^2 \right\rangle}{\left\langle \Pi \right\rangle^2} \right) }
 \end{equation}
 and their errors are, respectively
 \begin{equation}
{\rm var}(\mu) = \dfrac{4}{\left\langle \Pi \right\rangle^2} {\rm var}(\left\langle \Pi \right\rangle) + \dfrac{1}{\left\langle \Pi^2 \right\rangle} {\rm var}(\sqrt{\left\langle \Pi^2 \right\rangle)}
 \end{equation}
 \begin{equation}
{\rm var}(\sigma) = \dfrac{1}{\sigma^2} \left(  \dfrac{{\rm var}(\left\langle \Pi \right\rangle)}{\left\langle \Pi \right\rangle^2} + \dfrac{1}{\left\langle \Pi^2 \right\rangle} {\rm var}(\sqrt{\left\langle \Pi^2 \right\rangle)}  \right) 
 \end{equation}
 
 From $\mu$ we can then compute the median fractional polarisation as 
  \begin{equation}
\Pi_m=\exp{(\mu)}
 \end{equation}

Note that for the HFI channels the leakage effect is negligible when estimating the polarised flux density of compact sources \citep[see][]{PLA15pccs2}, so there is no need to take it into account in the simulations.

\section{Results}
\label{sec:results} 

Our sample consists of 4697 sources as described in \ref{sec:data}. We perform stacking in the \textit{Planck} channels from 143 to 353 GHz, both in total intensity and polarisation. From the patches resulting from stacking we estimate $\left\langle S \right\rangle$,  $\left\langle P_0 \right\rangle$, $\left\langle \Pi \right\rangle$ and $\left\langle S^2 \right\rangle$,  $\left\langle P_0^2 \right\rangle$, $\sqrt{\left\langle \Pi^2 \right\rangle}$ and we compute the errors from the standard deviation of the residual background fluctuations,  as described in  Section \ref{sec:stacking}. The results are summarized in the left part of Table~\ref{tab:polfrac_ext}.

\subsection{$\left\langle \Pi  \right\rangle $ and $\left\langle \Pi^2  \right\rangle $ estimate}
\label{sec:pi_estimate}
To estimate $\left\langle \Pi  \right\rangle $ from the bias-uncorrected values we use the procedure described in Section \ref{sec:s_injection}, whose results are shown in Fig. \ref{fig:ex_res} and listed in Table \ref{tab:polfrac_ext}. The same approach has been applied to correct also the values obtained for $\sqrt{\left\langle \Pi^2 \right\rangle}$ (see Table \ref{tab:polfrac_ext} and Fig. \ref{fig:ex_res_pi2}). 

Due to the large errors the mean fractional polarisation at 143 GHz is not well constrained ($\left\langle \Pi \right\rangle = (3.52 \pm 2.48)$ per cent and $\sqrt{\left\langle \Pi^2 \right\rangle} = (7.31 \pm 3.89)$ per cent).

Fig. \ref{fig:polfrac} is a summary of the noise bias corrected results on $\left\langle \Pi \right\rangle$ (top panel) and $\sqrt{\left\langle \Pi^2 \right\rangle}$ (bottom panel) we obtain with stacking on the current sample, compared with those obtained by \cite{BON17}. We obtain $\left\langle \Pi \right\rangle$ of $(3.10 \pm 0.75)$ per cent and $(3.65 \pm 0.66)$ per cent at 217 and 353 GHz, respectively. 

Interestingly, very similar values were obtained from the analysis of the radio sample in \cite{BON17}, i.e. $(3.07 \pm 0.29)$ per cent for 217 GHz and $(3.52 \pm 1.20)$ per cent for 353 GHz. We cannot extract any value of $\Pi$ from the PCCS2 catalogue for the dusty sources: we verified that all the sources in the PCCS2  catalogue at 143, 217 and 353 GHz with reliable detection in polarisation have a counterpart at 30 GHz, therefore we can consider them radio sources.

The similar results inferred from the analysis of the dusty and radio samples might imply a common origin for the observed polarisation. In this scenario, polarisation might be mainly caused by the synchrotron emission from the bursts of star formation and/or nuclear activity of dusty galaxies, with negligible contribution from dust. However, dust thermal emission contributes significantly to the total  intensity at these frequencies, decreasing the mean fractional polarisation. Therefore, in order to recover the observed $\sim 3$ per cent mean fractional polarisation, the contribution from polarised synchrotron emission should be unfeasibly high. Indeed, at 217 GHz, synchrotron emission of local dusty galaxies  \citep[the kind of sources in our sample,][]{NEG13} is on average of a few percent\citep{PEPXVI}, and is completely negligible at 353 GHz. This argument implies that dust might contribute significantly to the observed polarisation in the dusty sample.

Above ~100 GHz AGN with weak radio emission are dominated by dust emission from the torus or the host galaxy, as in typical sources dominated by thermal emission.
Therefore, the observed polarisation is likely associated with scattering processes off the dust grains and dust magnetic fields. On the other hand, the radio sample analysed in \cite{BON17} consists mainly of Blazar sources. The latter are dominated by the jet emission at all frequencies, thus the observed polarisation is to be ascribed to partially ordered magnetic fields in the jet \citep[][and references therein]{TUC12}. This is expected to be the case when the shocks in the jet compress an initial random field (with B perpendicular to the jet axis) or sheared to lie in plane (with B parallel to the jet axis).

Since the physics that produces these polarisation signals is completely different we can only conclude that this close similarity in the $\left\langle \Pi \right\rangle$ values can be just a coincidence. 

As in \cite{BON17}, from the recovered values of $\left\langle \Pi \right\rangle$ and $\sqrt{\left\langle \Pi^2 \right\rangle}$ we compute the parameters $\mu$ and $\sigma$ for the log-normal distribution and from $\mu$ we compute the median fractional polarisation as $\Pi_m=\exp({\mu})$. The resulting values are summarized in right part of Table \ref{tab:polfrac_ext}. 

\subsection{Polarisation distribution}
\label{sec:pol_distro}
We also recover unbiased values for $\left\langle \Pi \right\rangle$ and $\sqrt{\left\langle \Pi^2 \right\rangle}$ at 353 GHz using two different distributions for the polarisation fraction (see Figs. \ref{fig:ex_res_cfr} and \ref{fig:ex_res_cfr_pi2}). The aim is to test how other distributions work when estimating these quantities. We test the Normal and the gamma distributions. In the first case we obtain $(5.21 \pm 1.00)$ per cent for $\left\langle \Pi \right\rangle$ and $(6.39 \pm 1.54)$ per cent for $\sqrt{\left\langle \Pi^2 \right\rangle}$. For the Normal case the following relations between the ${\rm E[x]=\left\langle \Pi \right\rangle}$ and ${\rm E[x^2]=\left\langle \Pi^2 \right\rangle}$  hold:  ${\rm E[x] = \mu}$ and ${\rm E[x^2] = \mu^2 + \sigma^2}$. By substituting the values we obtain a negative value for $\sigma^2$, which is not possible and allow us to discard a Normal like distribution for the fractional polarisation. The same argument can be applied to the gamma distribution, for which we find $(4.82 \pm 0.78)$ per cent for $\left\langle \Pi \right\rangle$ and $(6.37 \pm 1.11)$ per cent for $\sqrt{\left\langle \Pi^2 \right\rangle}$. In this case the relations are ${\rm E[x]=k\theta}$ and ${\rm E[x^2]=k\theta^2+k^2\theta^2}$, that give a negative value for the product $k\theta^2$ when substituting the values we find. Considering that for the gamma family of distributions it should be $k>0$, we can confidently discard also this broad family of distributions for the fractional polarisation.

\subsection{Source counts and CMB power spectrum contribution}
\label{sec:srccounts} 

\begin{figure}
 \centering
 \includegraphics[width=\columnwidth]{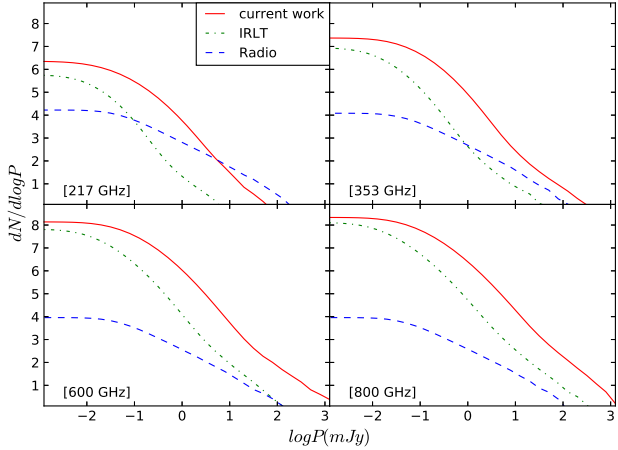}
 \caption{Source number counts estimated for our sample of sources (red solid line) and in the CORE predictions \citep{DEZ17} for the dusty (green dot-dashed line) and radio (blue dashed) sources. In the top panels we estimate the source number counts at 217 (left) and 353 (right) GHz with the results in Table \ref{tab:polfrac_ext}. In the lower panel we show the estimation at 600 (left) and 800 (right) GHz obtained assuming the same values for the log-normal parameters as for the 353 GHz case.}
 \label{fig:counts}
\end{figure}

\begin{figure}
 \centering
 \includegraphics[width=\columnwidth]{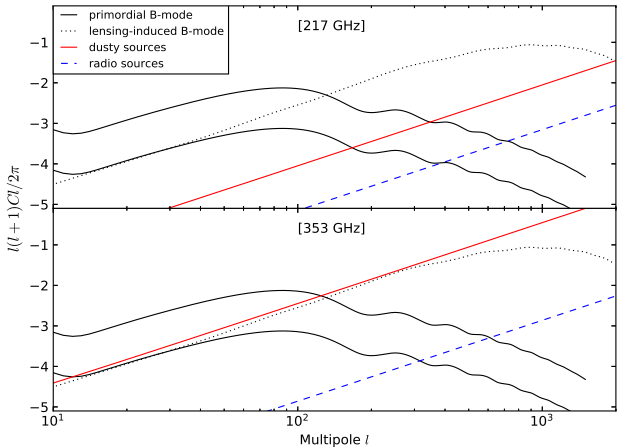}
 \caption{Power spectrum contribution from the dusty (red solid line) and radio (blue dashed line) sources to the power spectrum, compared with the primordial B-mode for $r=0.1, 0.01$, from top to bottom (black solid line) and the lensing-induced B-mode (black dotted line).}
 \label{fig:pw_cont}
\end{figure}

To estimate the source numbers counts in polarization, we proceed with simulations. At 217 and 353 GHz, we first simulate sources in total intensity according to the source number counts by \cite{NEG13}. Then we simulate the fractional polarization assuming a log-normal distribution with the parameters listed in Table \ref{tab:polfrac_ext}. Finally we randomly associate the simulated $\Pi$ value with the flux density in total intensity of the simulated sources. We repeat this process 10 times and we estimate the source number counts in polarization by averaging over these 10 sets of simulations.

We also estimate the source number counts in polarisation for 600 and 800 GHz (CORE frequencies) assuming the 353 GHz parameters for the log-normal distribution. The results are shown in Fig. \ref{fig:counts}: the red solid line are the source counts for our sample and they are compared with the dusty (green dashed line) and radio (blue dash-dotted line) sources predictions in the work by \cite{DEZ17}.

\begin{table}
 \centering
\begin{tabular}{ l | c | c | c | c  | c  | c | c | c | c | c | c | c | c || r } 
\hline
freq &  dusty sources  & Radio \\
$[GHz]$  & $[\mu K^{2}]$ & $[\mu K^{2}]$ \\
\hline
217 & 5.57e-08 & 4.38e-09 \\ 
353 & 2.22e-06 & 8.61e-09 \\ 
600 & 2.06e-03 & 5.66e-07 \\
800 & 5.10e-01 & 6.62e-05 \\
\hline
\end{tabular} 
 \caption{From left to right: frequency, $Cl$ contribution to the CMB power spectrum of our sample of dusty and of radio sources estimated from the CORE predictions \citep{DEZ17}.}
   \label{tab:pw_contr}
 \end{table}

From the source number counts in polarisation we estimate the contribution in polarisation to the CMB power spectrum, following \cite{DEZ96}:
\begin{equation}
Cl^P=g^2 \int_0^{P_c} P^2 \frac{dN}{dlogP} dlogP
\end{equation}
where g is the conversion factor from flux density to temperature units \citep{TEG96} and $P_c=0.5$ mJy has been chosen to be similar to the CORE detection limit \citep{DEZ17}. The values we obtain for 217, 353, 600 and 800 GHz are listed in Table \ref{tab:pw_contr}. In Fig. \ref{fig:pw_cont} we plot the $l(l+1)Cl/2\pi$ estimate for the dusty sources (red solid line) and for the radio sources in \cite{DEZ17} (blue dashed line) for the 217 (top panel) and 353 (bottom panel) GHz channels. In these two frequencies the contribution is comparable to the primordial B-mode for $r=0.1$ and $r=0.01$ (black solid lines) and the lensing-induced B-mode (black dotted line). At 217 the contribution from these kind of sources is negligible for the lensing-induced B-modes, but it becomes important for the primordial B mode at $l \sim 350$ (right after the second peak) or $l \sim 150$ (just after the first peak) for $r=0.1$ and $r=0.01$, respectively.

At 353 GHz the level of this contamination is the same as the one of lensing-induced B-modes. This means that it is even worse for the detection of the primordial B-mode, since the source contribution becomes important right after the first peak ($l \sim 100$) already for the $r=0.1$ case. 
We omit to show results from 600 and 800 GHz, as at these frequencies the contribution to the power spectrum is much higher than the one from the B-mode, as expected.

\section{Conclusions}
\label{sec:conc} 

The analysed sample of extragalactic dusty sources  (selected from the 857 GHz of the PCCS2 catalogue) shows polarisation properties similar to those characterizing radio sources \citep{MAS13,GAL17,BON17}. We measure polarization values of $\left\langle \Pi \right\rangle$ =$(3.10 \pm 0.75)$ per cent and $\left\langle \Pi \right\rangle$ = $(3.65 \pm 0.66)$ per cent  at 217 and 353 GHz respectively, and of $\sqrt{\left\langle \Pi^2 \right\rangle}$=$(7.38 \pm 1.32)$ per cent and $(6.87 \pm 1.35)$ per cent at 217 and 353 GHz, respectively.
Moreover, as for the radio sources, the fractional polarisation of extragalactic dusty sources follows a log-normal distribution. We find values for $\mu$ of $0.26 \pm 0.52$ and $0.66 \pm 0.41$ and for $\sigma$ of $1.32 \pm 0.23$ and $1.12 \pm 0.24$ at 217 and 353 GHz, respectively.
However, radio and dusty sources are dominated by different components at these frequencies, i.e. by jet synchrotron and dust emission respectively (see discussion in Section  \ref{sec:pi_estimate}). Therefore we conclude that the inferred similarities of polarisation properties are fortuitous.

We update with our new measurements the source number counts in polarisation at 217 and 353 GHz and compare them with the predictions for the CORE proposal \citep{DEZ17}.
Moreover we make prediction for the level of the expected contamination to the B-mode angular power spectrum. We also extrapolate the results at higher frequencies (600 and 800 GHz). We find that at 217 GHz the extragalactic dusty sources might be an important contaminant for the primordial B-mode, especially in the case of $r=0.01$ or lower. At 353 GHz their contribution is at the level of the lensing-induced B-mode. As expected, their importance increases with frequency and at 600 and 800 GHz their contribution to the angular power spectrum is much higher than the ones of the B-mode, both primordial and lensing-induced.

\section*{Acknowledgements}

L.B., J.G.N., F.A. and L.T. acknowledge financial support from the I+D 2015 project AYA2015-65887-P (MINECO/FEDER). J.G.N also acknowledges financial support from the Spanish MINECO for a 'Ramon y Cajal' fellowship (RYC-2013-13256). B.D.M. acknowledges support from the European Union's Horizon 2020 research and innovation programme under the Marie Sk{\l}odowska-Curie grant agreement No. 665778 via the Polish National Science Center grant Polonez UMO-2016/21/P/ST9/04025.


\bsp	
\label{lastpage}

\begin{thebibliography}{99} 

\bibitem[\protect\citeauthoryear{Andr{\'e} et al.}{2014}]{CORE}
Andr{\'e} P., Baccigalupi C. and Banday A., 2014, \jcap ,2, 006
%
\bibitem[\protect\citeauthoryear{Bennett et al.}{2003}]{BEN03} 
Bennett C.~L., et al., 2003, ApJ, 583, 1 
%
\bibitem[\protect\citeauthoryear{B{\'e}thermin et al.}{2012}]{BET12} 
B{\'e}thermin M., et al., 2012, A\&A, 542, A58 
%
\bibitem[\protect\citeauthoryear{BICEP2/Keck and Planck Collaborations et al.}{2015}]{BIC15} 
BICEP2/Keck and Planck Collaborations, et al., 2015, PhRvL, 114, 101301 
%
\bibitem[\protect\citeauthoryear{B17}{}]{BON17} 
Bonavera L., Gonz{\'a}lez-Nuevo J., Arg{\"u}eso F., Toffolatti L., 2017, MNRAS in press, arXiv:1703.09952
%
\bibitem[\protect\citeauthoryear{Choi \& Page}{2015}]{CHO15} 
Choi S.~K., Page L.~A., 2015, JCAP, 12, 020 
%
\bibitem[\protect\citeauthoryear{Curto et al.}{2013}]{CUR13} 
Curto A., Tucci M., Gonz{\'a}lez-Nuevo J., Toffolatti L., Mart{\'{\i}}nez-Gonz{\'a}lez E., Arg{\"u}eso F., Lapi A., L{\'o}pez-Caniego M., 2013, MNRAS, 432, 728 
%
\bibitem[\protect\citeauthoryear{De Zotti et al.}{1996}]{DEZ96} 
de Zotti G., Franceschini A., Toffolatti L., Mazzei P., Danese L., 1996, ApL\&C, 35, 289
%
\bibitem[\protect\citeauthoryear{De Zotti et al.}{2015}]{DEZ15} 
De Zotti G., et al., 2015, JCAP, 6, 018 
%
\bibitem[\protect\citeauthoryear{De Zotti et al.}{2016}]{DEZ17} 
De Zotti G., et al., 2017, JCAP accepted, arXiv:1609.07263 
%
\bibitem[\protect\citeauthoryear{Dole et al.}{2006}]{DOL06} 
Dole H., et al., 2006, A\&A, 451, 417 
%
\bibitem[\protect\citeauthoryear{Galluzzi et al.}{2017}]{GAL17} 
Galluzzi V., et al., 2017, MNRAS, 465, 4085
%
\bibitem[\protect\citeauthoryear{Greaves \& Holland}{2002}]{GRE02}
Greaves J.~S., Holland W.~S., 2002, AIPC, 609, 267 
%
\bibitem[\protect\citeauthoryear{Keck Array et al.}{2016}]{KEC16} 
Keck Array T., et al., 2016, arXiv, arXiv:1606.01968
%
\bibitem[\protect\citeauthoryear{Marsden et al.}{2009}]{MAR09} 
Marsden G., et al., 2009, ApJ, 707, 1729 
%
\bibitem[\protect\citeauthoryear{Massardi et al.}{2013}]{MAS13}
Massardi M., Burke-Spolaor S.~G., Murphy T.  et al., 2013, \mnras , 436, 2915-2928
%
\bibitem[\protect\citeauthoryear{Negrello et al.}{2013}]{NEG13} 
Negrello M., et al., 2013, MNRAS, 429, 1309
%
\bibitem[\protect\citeauthoryear{Planck Collaboration VII}{2011}]{ERCSC} 
Planck Collaboration VII, 2011, A\&A, 536, A7
%
\bibitem[\protect\citeauthoryear{Planck Collaboration XVI}{2011}]{PEPXVI} 
Planck Collaboration XVI, 2011, A\&A, 536, A16 
%
\bibitem[\protect\citeauthoryear{Planck Collaboration XIX}{2014}]{PLAisw} 
Planck Collaboration XIX, 2014, A\&A, 571, A19 
%
\bibitem[\protect\citeauthoryear{Planck Collaboration XIX}{2015}]{PIPXIX} 
Planck Collaboration XIX, 2015, A\&A, 576, A104 
%
\bibitem[\protect\citeauthoryear{Planck Collaboration I}{2016}]{PLA15gen} 
Planck Collaboration I, 2016, A\&A, 594, A1
%
\bibitem[\protect\citeauthoryear{Planck Collaboration XXVI}{2016}]{PLA15pccs2} 
Planck Collaboration XXVI, 2016, A\&A, 594, A26 
%
\bibitem[\protect\citeauthoryear{Planck Collaboration XXX}{2016}]{PIP30} 
Planck Collaboration XXX, 2016, A\&A, 586, A133
%
\bibitem[\protect\citeauthoryear{Remazeilles et al.}{2017}]{REM17} 
Remazeilles M., et al., 2017, arXiv, arXiv:1704.04501
%
\bibitem[\protect\citeauthoryear{Tegmark \& Efstathiou}{1996}]{TEG96} 
Tegmark M., Efstathiou G., 1996, MNRAS, 281, 1297 
%
\bibitem[\protect\citeauthoryear{Tucci \& Toffolatti}{2012}]{TUC12}
Tucci M. and Toffolatti L., 2012, Advances in Astronomy, 624987

\end{thebibliography}
\end{document}